\documentclass[12pt]{iopart}

\usepackage{graphicx}
\usepackage{makecell}
\usepackage{array}
\usepackage{multirow}
\usepackage{gensymb}
\newcolumntype{x}[1]{>{\centering\hspace{0pt}}p{#1}}
\usepackage{iopams}
\usepackage{longtable}
\usepackage{xcolor}
\usepackage{CJK}

\begin{document}

\title{Raman spectroscopy of K$_x$Co$_{2-y}$Se$_2$ single crystals near the ferromagnet-paramagnet transition}

\author{M. Opa\v{c}i\'c$^1$, N. Lazarevi\'c$^1$, M. M. Radonji\'c$^{2,3}$, M. \v{S}\'cepanovi\'c$^1$, Hyejin Ryu$^{4,5}$\footnote{Present address:Advanced Light Source, E. O. Lawrence Berkeley National Laboratory, Berkeley, California 94720, USA.}, Aifeng Wang$^4$, D. Tanaskovi\'c$^3$, C. Petrovic$^{4,5}$ and Z. V. Popovi\'c$^1$}
\address{$^1$ Center for Solid State Physics and New Materials, Institute of Physics Belgrade, University of Belgrade, Pregrevica 118, 11080 Belgrade, Serbia}
\address{$^2$ Center for Electronic Correlations and Magnetism, Theoretical Physics III, Institute of Physics, University of Augsburg,
D-86135 Augsburg, Germany}
\address{$^3$Scientific Computing Laboratory, Institute of Physics Belgrade, University of Belgrade, Pregrevica 118, 11080 Belgrade, Serbia}
\address{$^4$ Condensed Matter Physics and Materials Science Department, Brookhaven National Laboratory, Upton, New York 11973-5000, USA}
\address{$^5$ Department of Physics and Astronomy, Stony Brook University, Stony Brook, New York 11794-3800, USA}
\ead{nenad.lazarevic@ipb.ac.rs}

\date{\today}

\begin{abstract}
Polarized Raman scattering spectra of the K$_x$Co$_{2-y}$Se$_2$ single crystals reveal the presence of two phonon modes, assigned as of the A$_{1g}$ and B$_{1g}$ symmetry. Absence of additional modes excludes the possibility of vacancy ordering, unlike in K$_x$Fe$_{2-y}$Se$_2$. The ferromagnetic (FM) phase transition at $T_c\approx 74$ K leaves a clear fingerprint on the temperature dependence of the Raman mode energy and linewidth. For $T>T_c$ the temperature dependence looks conventional, driven by the thermal expansion and anharmonicity.
The Raman modes are rather broad due to the electron-phonon coupling increased by the disorder and spin fluctuation effects.
In the FM phase the phonon frequency of both modes increases, while an opposite trend is seen in their linewidth: the  A$_{1g}$ mode narrows in the FM phase, whereas the B$_{1g}$ mode broadens. We argue that the large asymmetry and anomalous frequency shift of the B$_{1g}$ mode is due to the coupling of spin fluctuations and vibration.
Our density functional theory (DFT) calculations for the phonon frequencies agree rather well with the Raman measurements, with some discrepancy being expected since
the DFT calculations neglect the spin fluctuations.
\end{abstract}

\maketitle

\section{Introduction}

In the last few years considerable attention was focused on the iron-based superconductors in an effort to gain deeper insight into their physical properties and to determine the origin of high-$T_c$ superconductivity \cite{Stewart_RevModPhys2011, Bao_ChinPhysLett2011, Liu_EPL2011, Ma_PRL2012}. Discovery of superconductivity in alkali-doped iron chalcogenides, together with its uniqueness among the iron based superconductors, challenged the physical picture of the superconducting mechanism in iron pnictides \cite{Dagotto_RevModPhys2013}. Absence of hole pockets even suggested the possibility for the different type of pairing mechanism \cite{Zhang_NatureMat2011}. Another striking feature in K$_x$Fe$_{2-y}$Se$_2$ was the presence of the intrinsic nano to mesoscale phase separation between an insulating phase and a metallic/superconducting phase \cite{Li_NatPhys2012,Lazarevic_PRB2012, Ding_NatComm2013, Louca_SciRep2013}. Insulating phase hosts antiferromagnetically, $\sqrt{5} \times \sqrt{5}$ ordered iron vacancies, whereas superconducting stripe-like phase is free of vacancies \cite{Li_NatPhys2012}.  Theoretical study of Huang \textit{et al.} \cite{Huang_PRB2013} revealed that proximity effects of the two phases result in the Fermi surface deformation due to interlayer hopping and, consequently, suppression of superconductivity. On the other hand, large antiferromagnetic order protects the superconductivity against interlayer hopping, thus explaining relatively high $T_c$ in K$_x$Fe$_{2-y}$Se$_2$ \cite{Huang_PRB2013}. However, the correlation between the two phases and its impact on superconductivity are still not fully understood.

Although the absolute values of resistivity are much smaller for the Ni-member of the K$_x$M$_{2-y}$Se$_2$ (M = transition metal) series than for the iron member, this material does not exhibit superconductivity down to 0.3 K \cite{Lei_JPCM2014}. As opposed to K$_x$Fe$_{2-y}$Se$_2$, vacancy ordering has not been observed in the K$_x$Ni$_{2-y}$Se$_2$ single crystal \cite{Lazarevic_PRB2013}. These materials, together with the Co- and Ni-doped K$_x$Fe$_{2-y}$Se$_2$ single crystals, have very rich structural, magnetic and transport phase diagrams. This opens a possibility for fine tuning of their physical properties by varying the sample composition \cite{Ryu_PRB2015, Ryu_PRB2015_2}.
First results obtained on K$_x$Co$_{2-y}$Se$_2$ single crystal revealed the ferromagnetic ordering below $T_c \sim$74 K, as well as the absence of the superconducting phase \cite{Yang_PRB2013}.

Raman spectroscopy is a valuable tool not only for measuring vibrational spectra, but also helps in the analysis of structural, electronic and magnetic properties, and phase transitions. There are several recent studies of the influence of the antiferromagnetic order, \cite{Um_PRB2012, Popovic_SSC2014} ferromagnetism, \cite{Eiter_PRB2014,Kirillov_PRB1995} and magnetic fluctuations \cite{Zhang_PRB2015} on the Raman spectra.

In this paper the Raman scattering study of K$_x$Co$_{2-y}$Se$_2$ single crystal (x=0.3, y=0.1), together with the lattice dynamics calculations of KCo$_2$Se$_2$, is presented. The polarized Raman scattering measurements were performed in the temperature range from 20 K up to 300 K. The observation of only two Raman active modes when measuring from the (001)-oriented samples suggests that K$_x$Co$_{2-y}$Se$_2$ single crystal has no ordered vacancies. The temperature dependence of the energy and linewidth of the observed Raman modes reveals a clear fingerprint of the phase transition. Large linewidth of the B$_{1g}$ mode and its Fano line shape indicate the importance of spin fluctuations.

The rest of the manuscript is organized as follows. Section II contains a brief description of the experimental and numerical methods, Section III are the results, and Section IV contains a discussion of the phonon frequencies and linewidths and their temperature dependencies. Section V summarizes the results.

\section{Experiment and numerical method}

Single crystals of K$_x$Co$_{2-y}$Se$_2$ were grown by the self-flux method, as described in Ref.~\cite{Lei_JPCM2014}.  The elemental analysis was performed using energy-dispersive x-ray spectroscopy (EDX) in a JEOL JSM-6500 scanning electron microscope. Raman scattering measurements were performed on freshly cleaved (001)-oriented samples with size up to 3$\times$3$\times$1 mm$^3$, using a TriVista 557 Raman system equipped with a nitrogen-cooled CCD detector, in backscattering micro-Raman configuration. The 514.5 nm line of an Ar$^+$/Kr$^+$ ion gas laser was used as an excitation source. A microscope objective with $50 \times$ magnification was used for focusing the laser beam. All measurements were carried out at low laser power, in order to minimize local heating of the sample. Low temperature measurements were performed using KONTI CryoVac continuous flow cryostat with 0.5 mm thick window. Spectra were corrected for the Bose factor.

The electronic structure of the ferromagnetic (FM) and paramagnetic (PM) phases is calculated within the density functional theory (DFT), and  the phonon frequencies at the $\Gamma$-point are obtained within the density functional perturbation theory (DFPT) \cite{Baroni_RevModPhys2001}. All calculations are performed using the QUANTUM ESPRESSO package \cite{Gianozzi_JPCM2009}. We have used Projector Augmented-Wave (PAW) pseudo-potentials with Perdew-Burke-Ernzerhof (PBE) exchange-correlation functional with nonlinear core correction and Gaussian smearing of 0.005 Ry. The electron wave-function and the density energy cutoffs are 40 Ry and 500 Ry, respectively. The Brillouin zone is sampled with 16$\times$16$\times$8 Monkhorst-Pack k-space mesh. The phonon frequencies were calculated with relaxed unit cell parameters and, for comparison, with the unit cell size taken from the experiments and the relaxed positions of only Se atoms. The forces acting on individual atoms in the relaxed configuration were smaller than $10^{-4}$ Ry/a.u.~and the pressure smaller than 0.5 kbar.

\section{Results}

\begin{figure}
\centering
\includegraphics[width = 0.4\textwidth]{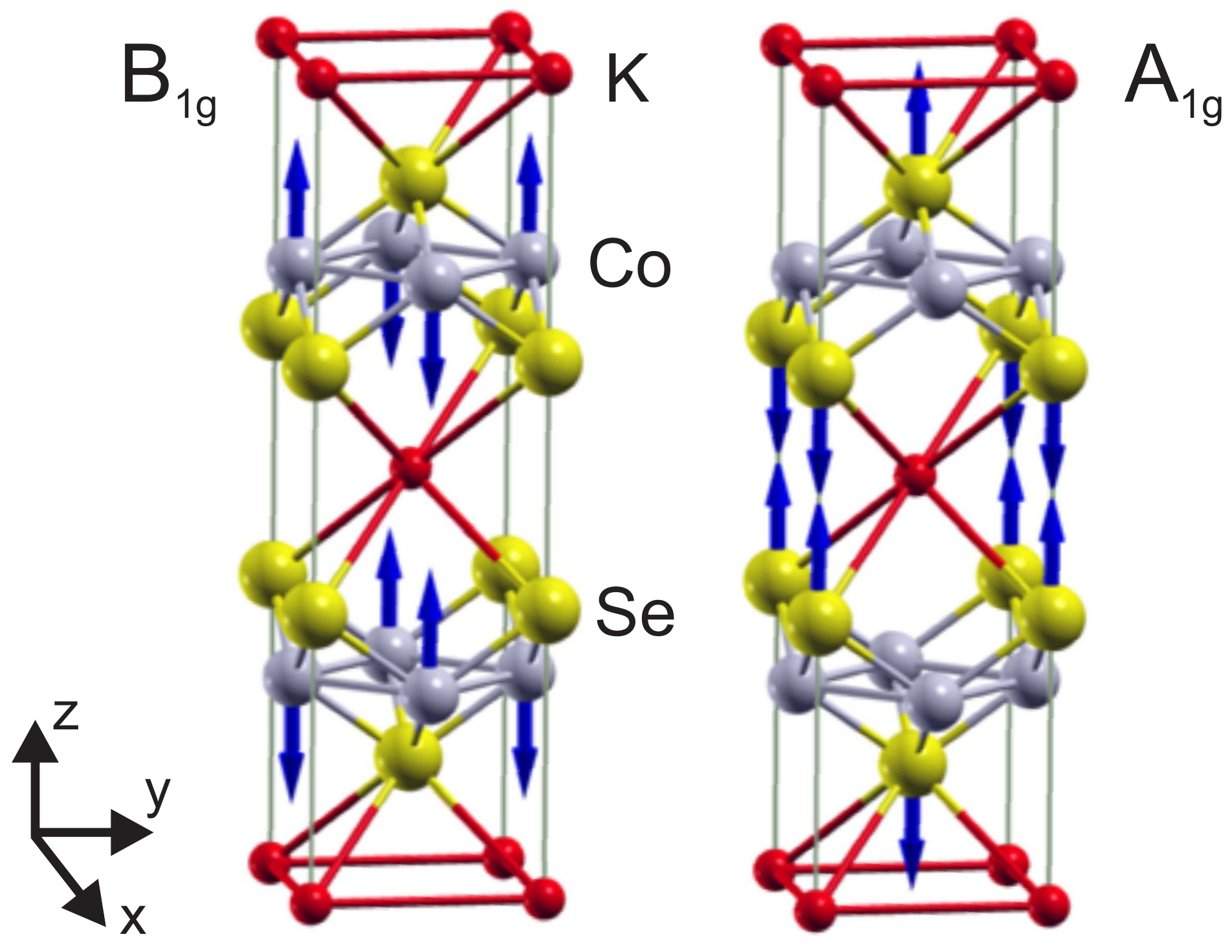}
\caption{(Color online) Unit cell of KCo$_2$Se$_2$ single crystal, together with the displacement patterns of the A$_{1g}$ and B$_{1g}$ Raman modes.}
\label{fig1}
\end{figure}

KCo$_2$Se$_2$ crystallizes in the tetragonal crystal structure of ThCr$_2$Si$_2$-type, $I4/mmm$ space group, which is shown in Figure \ref{fig1}. The experimental values of the unit cell parameters are $a=$3.864(2) $\AA$ and $c=$13.698(2) $\AA$ \cite{Guohe_JLCM1989}.
The potassium atoms are at $2a$:$(0,0,0)$, Co atoms at $4d$:$(0,\frac{1}{2},\frac{1}{4})$, and Se atoms at $4e$:$(0,0,z)$ Wyckoff positions, with the experimental value $z=0.347$.

KCo$_2$Se$_2$ single crystal consists of alternatively stacked K ions and CoSe layers, isostructural to the KFe$_2$Se$_2$ \cite{Guo_PRB2010}. Factor group analysis for the $I4/mmm$ space group yields a normal mode distribution at the Brillouin-zone center which is shown in Table~\ref{tab0}. According to the selection rules, when measuring from the (001)-plane of the sample, only two modes (A$_{1g}$ and B$_{1g}$) are expected to be observed in the Raman scattering experiment.
Displacement patterns of the experimentally observable Raman modes are illustrated in Fig.~\ref{fig1}. The A$_{1g}$ (B$_{1g}$) mode represents the vibrations of the Se (Co) ions along the $c$-axis, whereas the E$_g$ modes (which are not observable for our scattering configuration) involve the vibration of both Co and Se ions within the (001)-plane.

\begin{table*}[t]
\caption{Atomic types with their Wyckoff positions and the contribution of the each site to the $\Gamma$-point phonons, the Raman tensors and the selection rules for the K$_x$Co$_{2-y}$Se$_2$ single crystal ($I4/mmm$ space group).}
\label{tab0}
\centering
\resizebox{\linewidth}{!}{%
\begin{tabular}{c c c c}
\hline \hline
Atoms & Wyckoff positions & \multicolumn{2}{c} {Irreducible representations} \\

\cline{1-4} \\[-1.0em]

K & 2a & \multicolumn{2}{c} {A$_{2u}$+E$_u$} \\ [1mm]
Co & 4d & \multicolumn{2}{c} {A$_{2u}$+B$_{1g}$+E$_g$+E$_u$} \\ [1mm]
Se & 4e & \multicolumn{2}{c} {A$_{1g}$+A$_{2u}$+E$_g$+E$_u$}  \\ [1mm]

\multicolumn{4}{c}{Raman tensors} \\

$\hat{R}_{A_{1g}} = \left(\begin{array}{ccc} |a|\exp{i\varphi_a} & 0 & 0 \\ 0 & |a|\exp{i\varphi_a} & 0 \\ 0 & 0 & |b|\exp{i\varphi_b} \end{array}\right)$ & & $\hat{R}_{B_{1g}} = \left(\begin{array}{ccc} |c|\exp{i\varphi_c} & 0 & 0 \\ 0 & -|c|\exp{i\varphi_c} & 0 \\ 0 & 0 & 0 \end{array}\right)$ & \\[4mm]
$\hat{R}_{E_g} = \left(\begin{array}{ccc} 0 & 0 & |e|\exp{i\varphi_e} \\ 0 & 0 & 0 \\ |e|\exp{i\varphi_e} & 0 & 0 \end{array}\right)$ & & $\hat{R}_{E_g} = \left(\begin{array}{ccc} 0 & 0 & 0 \\ 0 & 0 & |f|\exp{i\varphi_f} \\ 0 & |f|\exp{i\varphi_f} & 0 \end{array}\right)$ & \\[1mm]

\multicolumn{4}{c}{Activity and selection rules} \\ [1mm]
\multicolumn{4}{c}{$\Gamma_{{Raman}} = A_{1g}(\alpha_{xx+yy}, \alpha_{zz})+B_{1g}(\alpha_{xx-yy})+2E_g(\alpha_{xz},\alpha_{yz})$} \\ [1mm]
\multicolumn{4}{c}{$\Gamma_{{IR}} = 2A_{2u}(\mathbf{E \| z})+2E_u (\mathbf{E\|x, E\|y})$} \\ [1mm]
\multicolumn{4}{c}{$\Gamma_{{acoustic}} = A_u+E_u$} \\ [1mm] \hline \hline

\end{tabular}}
\end{table*}

\begin{figure}
\centering
\includegraphics[width = 0.45\textwidth]{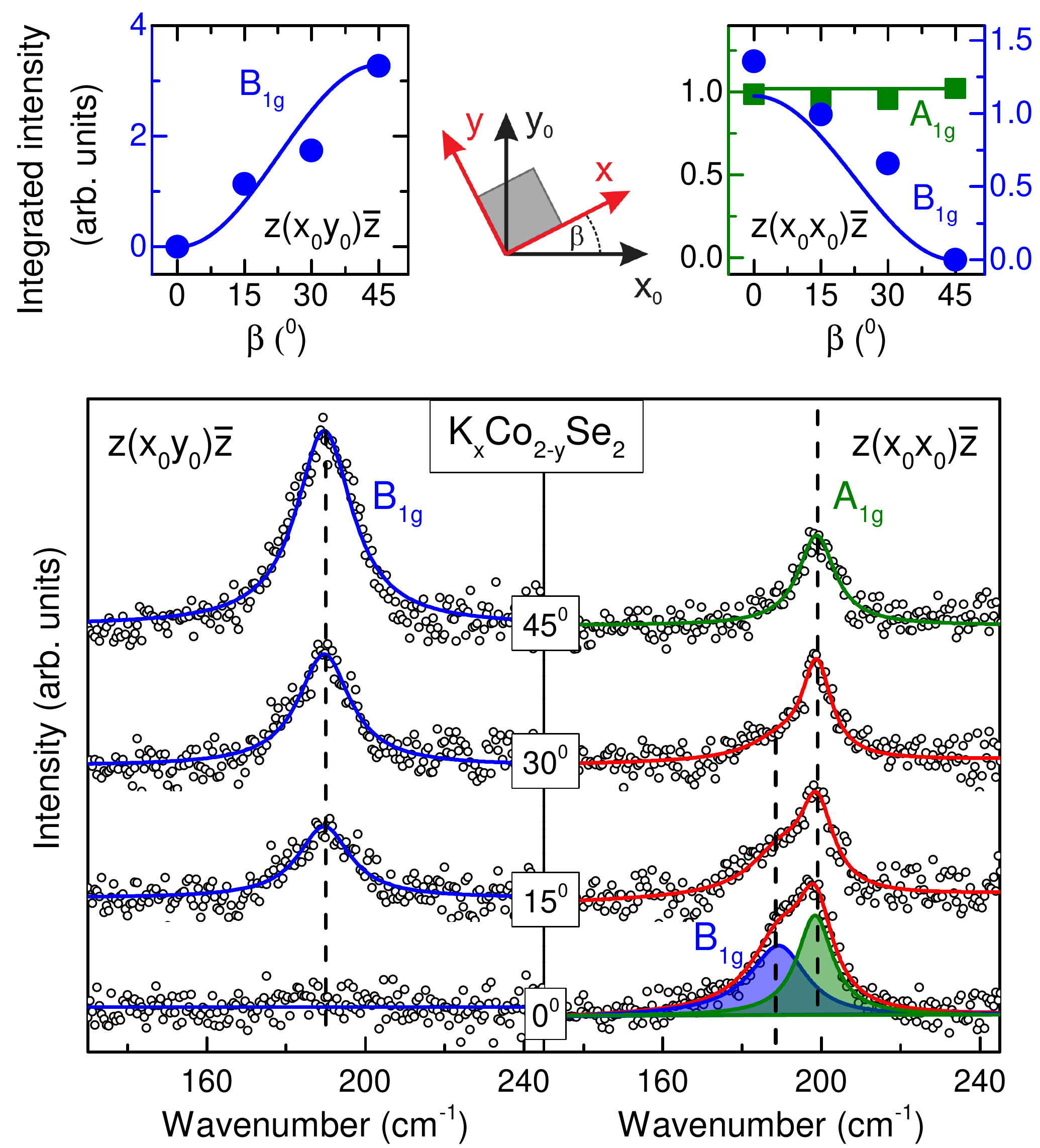}
\caption{(Color online) Upper panel: Integrated intensity of the observed Raman modes as a function of the crystal orientation with respect to the laboratory axes $\mathbf{x_0}$ and $\mathbf{y_0}$. In order to estimate the intensity of the modes, phonon at 198 cm$^{-1}$ was fitted with Lorentzian, whereas asymmetric Raman mode appearing at 187 cm$^{-1}$ was fitted with Fano line shape. Lower panel: Raman scattering spectra of K$_x$Co$_{2-y}$Se$_2$ single crystal measured at room temperature, in various sample orientations ($\mathbf{x}=[100], \mathbf{y}=[010]$).}
\label{fig2}
\end{figure}

Figure~\ref{fig2} shows polarized Raman scattering spectra of K$_x$Co$_{2-y}$Se$_2$ single crystal, measured from the (001)-plane of the sample at room temperature, in different sample orientations. Only two modes, at about 187 and 198 cm$^{-1}$, are observed, which is in agreement with the selection rules for (001)-oriented samples. In some iron-chalcogenide compounds, the appearance of additional Raman active modes due to the iron vacancy ordering and, consequently, symmetry lowering, has been observed \cite{Lazarevic_PRB2012, Lazarevic_PRB2011}. Absence of additional phonon modes in Fig. 2 suggests that in K$_x$Co$_{2-y}$Se$_2$ single crystals vacancy ordering does not occur in our samples.

Selection rules imply that the A$_{1g}$ mode may be observed for any sample orientation, provided that the polarization vector of the incident light $\mathbf{e}_i$ is parallel to the scattered light polarization vector $\mathbf{e}_s$, whereas it vanishes if these vectors are perpendicular. On the other hand, the intensity of the B$_{1g}$ mode strongly depends on the sample orientation ($I_{B_{1g}} \sim |c|^2 \cos^2 (\theta+2\beta)$, where $\theta=\angle(\mathbf{e}_i,\mathbf{e}_s)$ and $\beta = \angle(\mathbf{e}_i,\mathbf{x})$) \cite{Lazarevic_PRB2012}. This implies that, in parallel polarization configuration ($\theta=0^{\circ}$), the intensity of the B$_{1g}$ mode is maximal when the sample is oriented so that $\mathbf{e}_i$$\|$$\mathbf{x}$, gradually decreases with increasing $\beta$ and finally vanishes for $\beta=45^{\circ}$. In crossed polarization configuration ($\theta=90^{\circ}$), B$_{1g}$ mode intensity decreases from its maximal value for $\beta=45^{\circ}$ to zero, which reaches when $\beta=0^{\circ}$. From Fig.~\ref{fig2} it can be seen that the intensity of the Raman mode at about 187 cm$^{-1}$ coincides with theoretically predicted behavior for the B$_{1g}$ mode; thereby, this phonon mode is assigned accordingly. Phonon mode at $\sim$198 cm$^{-1}$, which is present in Raman spectra only for parallel polarization configuration ($\theta=0^{\circ}$) and whose intensity is independent on the sample orientation, can be assigned as the A$_{1g}$ mode. The intensity ratio of the two Raman modes can be obtained from the spectrum measured in ($\theta=0^{\circ}, \beta=0^{\circ}$) scattering geometry as $I_{B_{1g}}/I_{A_{1g}} \approx 1.38$. Having in mind that the A$_{1g}$ mode intensity is given by \cite{Lazarevic_PRB2012} $I_{A_{1g}} \sim |a|^2 \cos^2 \theta$, the ratio of the appropriate Raman tensor components can be estimated as $|c|/|a| \approx 1.17$.

\begin{table}[b]
\caption{Optimized lattice constants and internal coordinate $z_{_{Se}}$ in the FM and PM solution. Next two rows give the relaxed $z_{_{Se}}$ when the unit cell size is taken from the experiment, and the last row contains the atomic positions from the experiment \cite{Guohe_JLCM1989}}.
\label{table_unitcell}
\centering
\begin{tabular}{c c c c }
\hline  \hline
 & $a$ (\AA) &  $c$ (\AA)  &   $z_{_{Se}}$   \\

\cline{1-4} \\[-1.0em]

FM$^{rel}$   & 3.893 & 13.269 & 0.350  \\ [1mm]
PM$^{rel}$   & 3.766 & 13.851 & 0.368   \\ [1mm]
FM$^{fixed}$ & 3.864 & 13.698 & 0.3486  \\ [1mm]
PM$^{fixed}$ & 3.864 & 13.698 & 0.3496   \\ [1mm]
Exper.       & 3.864 & 13.698 & 0.347   \\ [1mm] \hline \hline

\end{tabular}
\end{table}

\begin{figure}
\centering
\includegraphics[width = 0.45\textwidth]{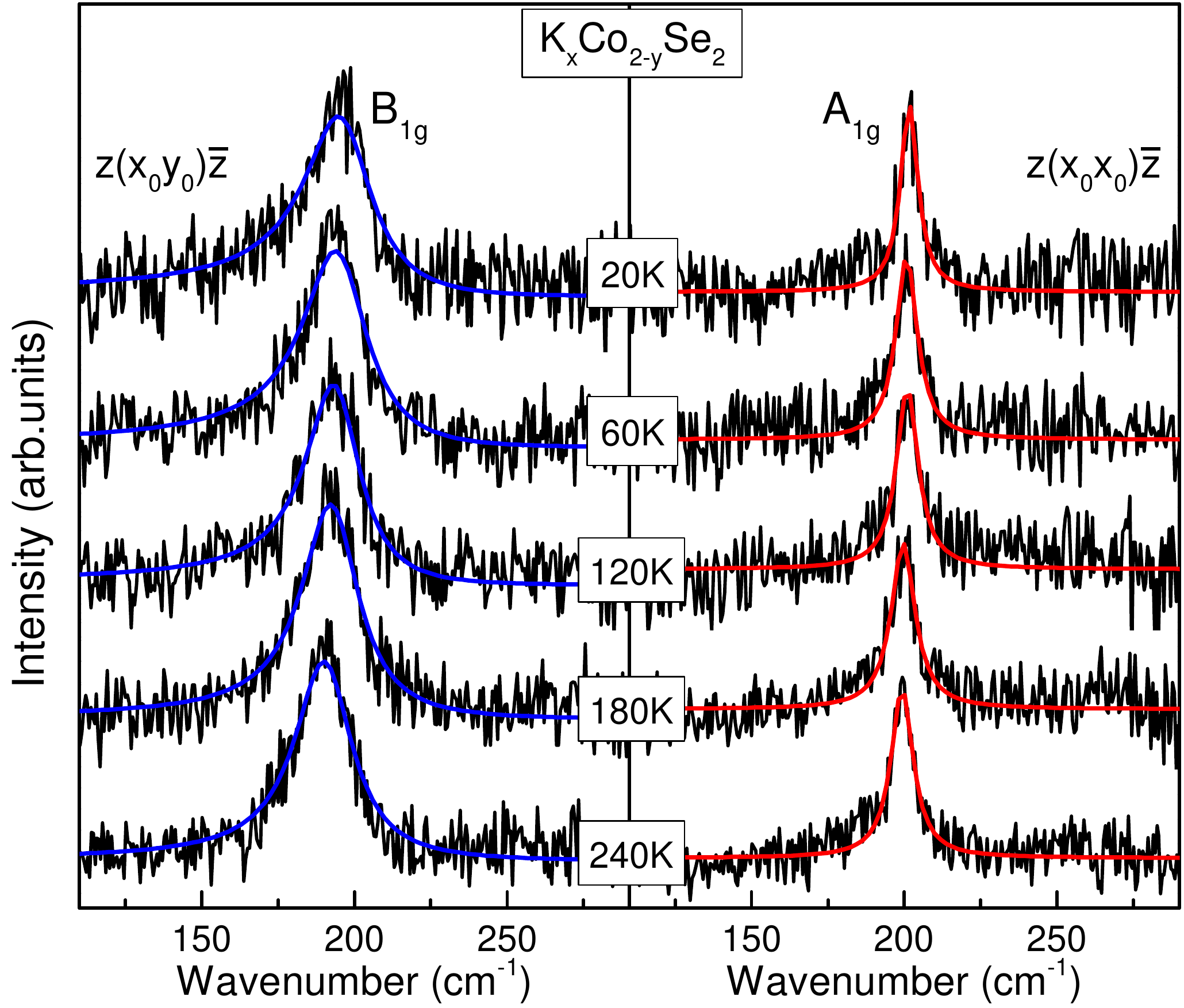}
\caption{(Color online) Temperature dependent Raman spectra of K$_x$Co$_{2-y}$Se$_2$ single crystal in parallel (left panel) and crossed (right panel) polarization configuration ($\mathbf{x_0}=\frac{1}{\sqrt{2}}[110], \mathbf{y_0} = \frac{1}{\sqrt{2}}[\bar{1}10]$). The solid lines represent fits of the experimental spectra with the Lorentzian (A$_{1g}$ mode) and the Fano profile (B$_{1g}$ mode).}
\label{fig3}
\end{figure}

\begin{table*}[t]
\caption{The experimental phonon energies measured at 20 K in the FM phase and the extrapolated value to 0 K from the PM phase (see the text). The phonon frequencies at the $\Gamma$ point are calculated with fully relaxed atomic positions. The frequencies obtained with relaxed only internal coordinate are given in parenthesis. }
\label{tab1}
\centering
\resizebox{1.0\textwidth}{!}{%
\begin{tabular}{c c c c c c c}
\hline \hline
Symmetry & Activity &  \begin{tabular}{c} Experiment \\ FM (cm$^{-1}$) \end{tabular}  & \begin{tabular}{c} Experiment \\ PM (cm$^{-1}$) \end{tabular} & \begin{tabular}{c} Calculation \\ FM (cm$^{-1}$) \end{tabular} & \begin{tabular}{c} Calculation \\ PM (cm$^{-1}$)   \end{tabular} & \begin{tabular}{c} Main atomic \\ displacements  \end{tabular} \\

\cline{1-7} \\[-1.0em]

A$_{1g}$ & Raman & 201.9 & 201.3 & 199.5 (193.2) & 212.6 (193.1) & Se(z) \\ [1mm]
B$_{1g}$ & Raman & 195.3 & 194.2 & 171.2 (172.7) & 176.6 (168.1) & Co(z)  \\ [1mm]
E$_g^1$ & Raman & & & 93.1 (100.7) & 92.7 (99.0) & Co(xy), Se(xy)  \\ [1mm]
E$_g^2$ & Raman & & & 237.9 (237.6) & 257.2 (235.6) & Co(xy), Se(xy)  \\ [1mm]
A$_{2u}^1$ & IR & & & 115.1 (99.0) & 113.7 (102.9) & K(z), Se(-z)  \\ [1mm]
A$_{2u}^2$ & IR & & & 246.7 (241.4) & 250.9 (241.4) & Co(z), K(-z)  \\ [1mm]
E$_u^1$ & IR & & & 97.9 (95.0) & 100.1 (95.0) & K(xy)  \\ [1mm]
E$_u^2$ & IR & & & 239.0 (229.7) & 231.0 (229.9) & Co(xy), Se(-xy)  \\ [1mm] \hline \hline

\end{tabular}}
\end{table*}

\begin{figure*}
\begin{center}
\includegraphics[width = 0.9\textwidth]{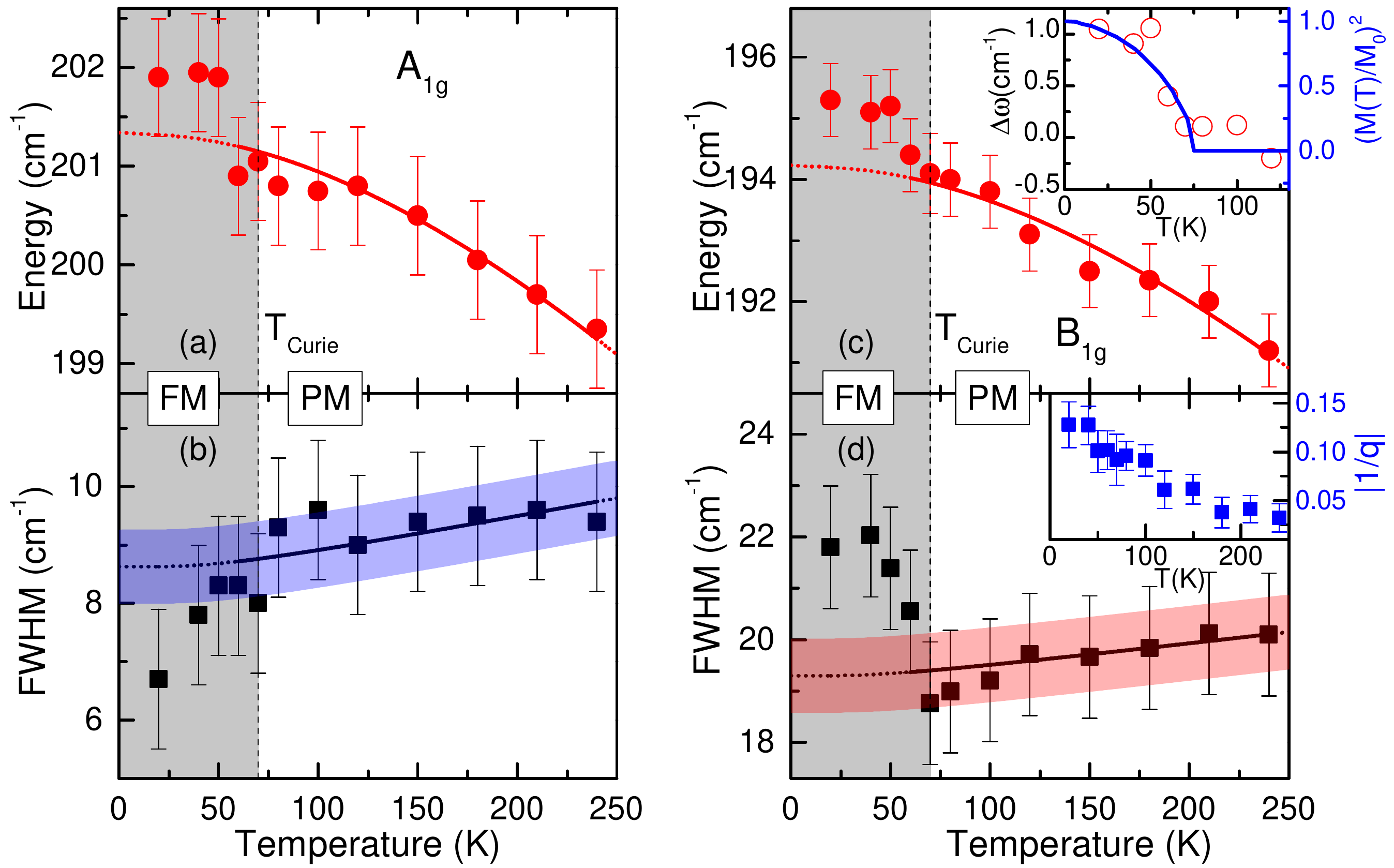}
\caption{(Color online) Temperature dependence of the energy and linewidth for the A$_{1g}$ (a, b) and B$_{1g}$ (c, d) Raman modes of K$_x$Co$_{2-y}$Se$_2$ single crystal. Solid lines are a theoretical fit (see the text) and the dotted lines are the extrapolation to the FM phase. Upper inset: Temperature dependence of the B$_{1g}$ mode frequency, compared with $(M(T)/M(0))^2$ curve. Lower inset: Measure of the electron-mediated photon-phonon coupling ($1/q$) of the B$_{1g}$ mode as a function of temperature.}
\label{fig4}
\end{center}
\end{figure*}

The experimentally determined frequencies are compared with those obtained with DFT numerical calculations. The experimental lattice constants \cite{Guohe_JLCM1989} are shown in Table II, together with their values from the DFT calculation which relaxes or keeps fixed the unit cell size. The DFPT phonon frequencies obtained using the fully relaxed atomic positions
in both FM and PM phases are given in Table III, with the corresponding values obtained with the fixed unit cell size and relaxed only fractional coordinate $z_{_{Se}}$ given in the parenthesis.
The equilibrium atomic positions in the FM solution are given by $a=3.893$ \AA, $c=13.269$ \AA, and $z_{_{Se}}=0.350$. The corresponding phonon frequencies are 199.5 $\mathrm{cm}^{-1}$ for A$_{1g}$ mode and 171.2 $\mathrm{cm}^{-1}$ for B$_{1g}$ mode. When we enforce the PM solution, we obtain $a=3.766$ \AA, $c=13.851$ \AA, and $z_{_{Se}}=0.368$, and 212.6 $\mathrm{cm}^{-1}$, 176.6 $\mathrm{cm}^{-1}$ for the frequencies of the A$_{1g}$ and B$_{1g}$ mode, respectively. These values agree rather well with the experimental data, and agree with recently published numerical results \cite{Wdowik_JPCM2015, Note_typo}. They can be used to confirm the experimental assignment of the modes, but cannot resolve subtle changes of the phonon frequencies near the FM-PM transition. This level of discrepancy is expected for metallic materials with magnetic ordering since the DFT calculations neglect spin fluctuations, as discussed in some detail in the next Section (see also Ref.~\cite{Zhang_PRB2015}). A rather large difference between the calculated frequencies in the two phases is due the relatively large change in the unit cell size. This difference between the unit cell sizes in the FM and PM phases is overestimated in the calculation which neglects spin fluctuations. For comparison, we also calculated the frequencies keeping the experimental values of the unit cell size, and relaxing just the coordinate $z_{_{Se}}$ of the Se atoms, which is often done in the case of iron based superconductors and related compounds \cite{Zhang_PRB2015}. This gives $z_{_{Se}}=0.3486$ in the FM solution and $z_{_{Se}}=0.3496$  in the PM solution, while the change in the phonon frequencies between the two solutions is much smaller, see Table III and a discussion in Section IV.

Polarized Raman scattering spectra of K$_x$Co$_{2-y}$Se$_2$ single crystals, measured at various temperatures from the (001)-plane of the sample, are presented in Figure~\ref{fig3}. Orientation of the sample is chosen so that each of the observable modes appears in different polarization configuration. A pronounced feature in the spectra is an asymmetric Fano profile of the B$_{1g}$ mode, persisting down to low temperatures, as well as its large linewidth compared to isostructural K$_x$Fe$_{2-y}$Se$_2$ \cite{Lazarevic_PRB2012, Opacic_JPCM2015}. This feature should by mainly due to the spin fluctuations influencing the B$_{1g}$ vibrational mode which modulates the distances between the magnetic Co atoms. A detailed discussion of the frequency and linewidth temperature dependence is given in the next Section.

\section{Discussion}

There are several factors that affect the phonon frequencies (energies) and linewidths, and their changes across the FM-PM transition. In general, the temperature dependence of the phonon frequency of the mode $i$, $\omega_i(T)$, is influenced by thermal expansion and magnetostriction, anharmonicity effects, electron-phonon and magnetic exchange interaction (spin-phonon coupling) \cite{Granado_PRB1999,Gupta_PRB2002}
\begin{eqnarray}\label{T_dependance}
\omega_i(T) - \omega_i(T_0) &=& \Delta \omega_i(T) = (\Delta \omega_i)_{latt} + (\Delta \omega_i)_{anh} \nonumber \\
&+& (\Delta \omega_i)_{el-ph} + (\Delta \omega_i)_{sp-ph}.
\end{eqnarray}
The first term is the frequency shift due to the change of the unit cell size caused by the thermal effects and magnetostriction. $(\Delta \omega_i)_{anh}$ is the anharmonic frequency shift. $(\Delta \omega_i)_{el-ph}$ appears due to the change in the electron-phonon interaction primarily influenced by changes in the electronic spectrum near the Fermi level, and $(\Delta \omega_i)_{sp-ph}$ is the spin-phonon contribution caused by the modulation of exchange interaction by lattice vibrations.

In our case of K$_x$Co$_{2-y}$Se$_2$, for temperatures above $T_c$, $\omega_i(T)$ decreases and $\Gamma_i(T)$ (full width at half-maximum, FWHM) increases with increasing temperature for $A_{1g}$ and $B_{1g}$ modes, similar as in the Raman spectra of many other materials. However, they show anomalous behavior near $T_c$, see Figure \ref{fig4}. In the following, we analyze $\omega_i(T)$ and $\Gamma_i(T)$ more closely.

\subsection{Phonon frequencies}

The frequencies of the A$_{1g}$ and B$_{1g}$ modes change by less than 2 percent in the temperature range between 20 K and 250 K. The red solid lines in Figs.~\ref{fig4}(a),(c)
represent the fits of the phonon energy temperature dependence (see below), following the frequencies of the two modes in the high-temperature PM phase.
The red dotted line is the extrapolation to $T=0$. For $T>T_c$, the temperature dependence of the frequency looks conventional for both modes: the frequency decreases with increasing temperature. This behavior is expected both due to the thermal expansion and the anharmonicity. These two effects can be standardly analyzed as follows.

The temperature dependent frequency of the vibrational mode $i$ is given by
\begin{equation}
\omega_i(T)=\omega_{0,i}+\Delta_i(T),
\label{A1}
\end{equation}
where $\omega_{0,i}$ denotes the temperature independent term and $\Delta_i(T)$ can be decomposed as \cite{Menendez_PRB1984, Haro_PRB1986,Eiter_PRB2014}
\begin{equation}
\Delta_i(T)=\Delta_i^V+\Delta_i^A.
\label{A2}
\end{equation}
$\Delta_i^V$ describes a change of the Raman mode energy as a consequence of the lattice thermal expansion and can be expressed with \cite{Menendez_PRB1984}
\begin{equation}
\Delta_i^V = \omega_{0,i}\left(e^{-3\gamma_i \int_{0}^{T} \alpha(T')dT'}-1\right),
\label{A3}
\end{equation}
where $\gamma_i$ is the Gr\"{u}neisen parameter of the Raman mode $i$ and $\alpha(T)$ is the thermal expansion coefficient of a considered single crystal. $\Delta_i^A$ represents the anharmonic contribution to the Raman mode energy.
If we assume, for simplicity, that anharmonic effects are described by three-phonon processes, this term is given by \cite{Menendez_PRB1984,Rahlenbeck_PRB2009}
\begin{equation}\label{freq}
\Delta_i^A = -C_i \left(1+\frac{2 \lambda_{p-p,i}}{e^{\hbar \omega_{0,i}/2k_BT}-1}\right),
\end{equation}
where $C$ is the anharmonic constant and $\lambda_{p-p,i}$ is a fitting parameter which describes the phonon-phonon coupling, including the nonsymmetric phonon decay processes.

The relative importance of the thermal expansion and anharmonicity to frequency changes is, to the best of our knowledge, not yet firmly established for pnictides and chalcogenides. In several cases \cite{Um_PRB2012,Lazarevic_PRB2013} the anharmonic formula, Eq.~(\ref{freq}), is used for the $\omega(T)$ fit. We follow here the arguments from Refs.~\cite{Gnezdilov_PRB2013, Eiter_PRB2014,Opacic_JPCM2015} that $\omega(T)$ is dominated by the thermal expansion.
To the best of our knowledge, the thermal expansion coefficient $\alpha(T)$ of K$_x$Co$_{2-y}$Se$_2$ single crystal is unknown. For estimating the lattice thermal expansion contribution to the phonon energy change, the coefficient $\alpha(T)$ for FeSe, given in
Ref.~\cite{Bohmer_PRB2013}, is used. The best fit shown in our Fig.~4 is obtained with  $\omega_{0,A_{1g}} = 201.3$ cm$^{-1}$, $\gamma_{A_{1g}} =1.23$ and $\omega_{0,B_{1g}} = 194.2$ cm$^{-1}$, $\gamma_{B_{1g}} =1.7$.

There exists a shift in phonon frequencies as the temperature is lowered below $T_c$. This shift does not show clear discontinuity (as well as the corresponding shift in the linewidths) and no additional modes are registered in the Raman spectra, which suggest that the FM-PM transition is continuous, without structural changes. There are several causes of the sudden frequency change as the sample gets magnetized. It can change due to the magnetostriction, modulation of the magnetic exchange by lattice vibrations (spin-phonon coupling), and due to the changes in the electron-phonon interaction due to spin polarization and changes in the electronic spectrum.

The effect of spin-phonon interactions, caused by the modulation of magnetic exchange interaction by lattice vibrations, may be quantitatively examined within the framework developed in Ref.~\cite{Granado_PRB1999} for insulating magnets, and recently applied also to several itinerant ferromagnets \cite{Kumar_PRB2014, Iliev_PRB2007, Laverdiere_PRB2006, Kumar_SSC2014}. In this model, the shift of the Raman mode energy due to the spin-phonon interaction is proportional to the spin-spin correlation function $\langle S_i|S_j \rangle$ between nearest magnetic ions. This term should have the same temperature dependence as $(M(T)/M_0)^2$, where $M(T)$ is the magnetization per magnetic ion at a temperature $T$ and $M_0$ is the saturation magnetization,
\begin{equation}
\Delta \omega(T) = \omega_{{exp}}(T)-\omega_{{fit}}(T) \propto \pm \left(\frac{M(T)}{M_0}\right)^2,
\label{eq6}
\end{equation}
where $\omega_{fit}(T)$ is the extrapolation from the high-temperature data.
This model does not predict the sign of the phonon energy shift - softening or hardening.
From the inset in Fig.~\ref{fig4}(c) it can be seen that the B$_{1g}$ mode energy renormalization scales well with the $(M(T)/M_0)^2$ curve. However, the effect of the magnetostriction (change of the unit cell size due to the magnetization) cannot be excluded based just on this plot, especially since the A$_{1g}$ mode corresponding to the vibrations of nonmagnetic Se ions also shows a similar shift in frequency.

The DFT calculations can give us some guidance for understanding of the changes of the phonon frequencies and linewidths, but one has to be aware of its limitations. The DFT calculations (see Table II) give a rather large magnetostriction, i.e.~rather large change in the size of the unit cell between the FM and PM phases ($a$ changes by 3.2\% and $c$ by 4.3\%). This leads to very large changes in the phonon frequencies, see Table III. The calculated frequencies are lower in the FM phase, as opposed to the experimental data. This already points to the limitations of the DFT calculations, which is expected near the phase transition. Similar conclusion is also present in Ref.~\cite{Zhang_PRB2015}. The DFT ignores spin fluctuations which often leads to quantitative discrepancy in various physical quantities \cite{Yin_NatureMat2011} and, in some cases, even predicts wrong phases. In the case of K$_x$Co$_{2-y}$Se$_2$, the DFT calculations correctly predict the FM ground state, but the calculated magnetic moment $m=0.947 \mu_B$ is much larger than the experimental value $m\approx 0.72 \mu_B$ \cite{Yang_PRB2013}. This already shows the importance of correlations and quantum fluctuations which are neglected within the DFT. Strong correlation effects can be captured using screened hybrid functional \cite{Yin_PRX2013} or within the dynamical mean field theory combined with DFT (LDA+DMFT) \cite{Haule_PRL2008}, which is beyond our present work.

Since the magnetostriction effects are overestimated in the DFT calculations with relaxed unit cell size, we repeated the DFT (DFPT) calculations keeping the experimental value for the unit cell size and relaxing only the fractional coordinate (positions of the Se atoms). This is often done in the literature on iron based superconductors and related compounds \cite{Zhang_PRB2015}. Our calculated frequencies are given in the parenthesis in Table III. We see that the frequency changes between the two phases are small, in better agreement with the experiment.

\subsection{Phonon linewidths}

The phonon linewidths of the A$_{1g}$ and B$_{1g}$ modes are very large, $\Gamma_{i,A_{1g}} \sim 10$ $\mathrm{cm}^{-1}$ and  $\Gamma_{i,B_{1g}} \sim 20$ $\mathrm{cm}^{-1}$, which implies the importance of disorder (impurities, nonstoichiometry, lattice imperfections) in measured samples. In general, the broadening of the phonon lines can be a consequence of the electron-phonon interaction, disorder, spin fluctuations and anharmonicity effects. The temperature dependence of the linewidth in the PM phase is, however, very weak, which indicates that the anharmonicity effects are small. The DFT calculation of the linewidth is usually based on the Allen's formula, \cite{Allen_PRB1972}
$\Gamma_{{\bf q},i} =  \pi N(E_F)\lambda_{{\bf q},i} \omega_{{\bf q},i}^2$.
Here, $N(E_F)$ is the density of states (DOS) at the Fermi level, $\lambda_{{\bf q},i}$ is the electron-phonon coupling constant, and $\omega_{{\bf q},i}^2$ is the phonon frequency of the mode $i$ and wavevector ${\bf q}$. A straightforward implementation of Allen's formula in the ${\bf q} \rightarrow 0$ limit corresponding to the $\Gamma$ point is, however, unjustified, as explained for example in Refs.~\cite{Cappelluti_PRB2006,Calandra_PRB2005}. In addition, structural disorder and impurities break the conservation of the momentum, which means that phonons with finite wave vectors also contribute to the Raman scattering spectra. The standard DFT calculation for the Brillouin zone averaged electron-phonon coupling constant $\lambda$ gives too small value to explain the large width of the Raman lines in pnictides and chalcogenides,\cite{Rahlenbeck_PRB2009} and several other metallic systems like MgB$_2$ \cite{Cappelluti_PRB2006} and fullerides \cite{Aksenov_PRB1998}. A correct estimate of the phonon linewidth can be obtained
only by explicitly taking into account the disorder and electron scattering which enhances the electron-phonon interaction,\cite{Cappelluti_PRB2006,Aksenov_PRB1998} which is beyond the standard DFT approach and scope of the present work.

The Raman mode linewidth is not directly affected by the lattice thermal expansion.
Assuming that the three-phonon processes represent leading temperature dependent term in the paramagnetic phase,  full width at half-maximum, $\Gamma_i(T)$, is given by
\begin{equation}
\Gamma_i(T) = \Gamma_{0,i} \left(1+\frac{2\lambda_{p-p,i}}{e^{\hbar \omega_{0,i}/2k_BT}-1}\right)+A_i.
\label{A5}
\end{equation}
\noindent The first term represents the anharmonicity induced effects, where $\Gamma_{0,i}$ is the anharmonic constant. The second term $A_i$ includes the contributions from other scattering channels, i.e. structural disorder and/or coupling of phonons with other elementary excitations, like particle-hole and spin excitations. These effects, typically, depend very weakly on temperature, but can become important near the phase transition. The best fit parameters are $\lambda_{p-p,i}=0.2$ for both modes, $A_{A_{1g}}=6.6$ cm$^{-1}$ and $A_{B_{1g}}=17.3$ cm$^{-1}$. The value $\Gamma_{0,i}=2$ cm$^{-1}$ is adopted from Ref.~\cite{Opacic_JPCM2015} for related compound K$_x$Fe$_{2-y}$Se$_2$, where the anharmonic effects dominate the temperature dependence. We see that $\lambda_{p-p,i}$ assume values much smaller than 1. Small and sometimes irregular changes in $\Gamma_i(T)$ are also observed in other materials whose Raman spectra are considered to be dominated by spin fluctuations \cite{Rahlenbeck_PRB2009,Zhang_PRB2015}. Therefore, we believe that a simple separation of $\Gamma_i(T)$ to the anharmonic and temperature independent term, which works well in many systems, is not appropriate for itinerant magnetic systems like K$_x$Co$_{2-y}$Se$_2$. We conclude that the spin fluctuations and electron-phonon coupling are likely to affect the linewidth even above $T_c$.

The electron-phonon interaction strength is proportional to the density of states at the Fermi level $N(E_F)$. Our DFT calculations for the DOS agree with those in Ref.~\cite{Bannikov_PhysicaB2012}. The calculated DOS in the FM phase, $N(E_F)=3.69$/eV, is smaller than, $N(E_F)=5.96$/eV, in the PM phase. (Though, in reality, it is possible that the DOS significantly differs from the one given by the DFT calculations due to the spin fluctuations and disorder effects.) Therefore, one expects that the phonon line is narrower in the FM phase than in the PM phase. This is indeed the case for the A$_{1g}$ mode, but the opposite is observed for the B$_{1g}$ mode.

It is also interesting to note that the B$_{1g}$ mode is much more asymmetric than the A$_{1g}$ mode and almost twice broader. These two observations are in striking similarity with the Raman spectra in the quasi-one-dimensional superconductor K$_2$Cr$_3$As$_3$ \cite{Zhang_PRB2015}. In this material the vibrational mode that modulates the distance between the magnetic Cr atoms also features large asymmetry and linewidth. In our case, the distances between
the magnetic Co ions are modulated by the vibrations of the B$_{1g}$ mode, see Fig.~1. This leads us to the conclusion that the anomalous features of the B$_{1g}$ mode are the consequence of spin fluctuations coupled to the electronic structure via lattice vibrations (in addition to the magnetostriction and spin polarization, which change the electronic spectrum near the Fermi level and, therefore, affect the electron-phonon interaction for both modes). 
It should be noted that similar anomalous properties of B$_{1g}$ phonon were  experimentally observed in cuprate high-temperature superconductor YBa$_2$Cu$_3$O$_7$ \cite{Ruf_PRB1988, Macfarlane_SSC1987}, and explained as a consequence of the out-of-phase nature of this mode which couples to oxygen-oxygen in-plane charge fluctuations \cite{Barisic_IJMPB1989,Devereaux_PRB1995,Kupcic_PRB2007} In the case of iron-based superconductors and related compounds, the chalcogen atoms and Fe (or Co) are not in the same plane and phonons of A$_{1g}$ symmetry can also directly couple with the electrons. A satisfactory agreement of theory and Raman experiments remains to be established \cite{Garcia-Martinez_PRB2013}.

The asymmetric B$_{1g}$ phonon line can be described by the Fano profile \cite{Iliev_PRB2005,Zhang_PRB2015,Kumar_PRB2014, Lazarevic_PRB2010}
\begin{equation}
I(\omega)=I_0 \frac{(\epsilon + q)^2}{1+\epsilon^2},
\label{eq1}
\end{equation}
where $\epsilon = 2(\omega - \omega_0)/\Gamma$, $\omega_0$ is the bare phonon frequency, $\Gamma$ is the linewidth. $I_0$ is a constant and $q$ is the Fano asymmetry parameter. It serves as a measure of a strength of the electron-phonon coupling: an increase in $|1/q|$ indicates an increase in the electron-phonon interaction, more precisely, an increase in the electron-mediated photon-phonon coupling function \cite{Devereaux_PRB1995,Garcia-Martinez_PRB2013}. From the inset of Fig.~\ref{fig4}(d) it can be seen that $|1/q|$ increases as the temperature is lowered and reaches the highest values around $T_c$, when the spin fluctuations are the strongest. Spin fluctuations increase the electron-phonon scattering, similar as does the disorder. Technically, the electronic Green function acquires an imaginary component of the self energy due to the spin fluctuations, and this implies the increase in the damping term in the phonon self-energy, as explained in, e.g. Ref.~\cite{Cappelluti_PRB2006}. This leads us to conclude that the spin fluctuations strongly enhance the electron-phonon interaction for the B$_{1g}$ vibrational mode affecting its frequency and linewidth near $T_c$.

\section{Conclusion}

In summary, the Raman scattering study of the K$_x$Co$_{2-y}$Se$_2$ $(x=0.3, y=0.1)$ single crystals and lattice dynamics calculations of the KCo$_2$Se$_2$, have been presented. Two out of four Raman active phonons are experimentally observed and assigned. The lack of any additional modes indicates the absence of vacancy ordering. The Raman spectra show sudden changes in the phonon energy and linewidth near the FM-PM phase transition. Above $T_c$ the energy and linewidth temperature dependence of the A$_{1g}$ and B$_{1g}$ modes look conventional, as expected from the thermal expansion and anharmonicity effects. The linewidth, though, has very weak temperature dependence even above $T_c$ which may the consequence of the proximity of the phase transition and spin fluctuations.  The B$_{1g}$ vibrational mode has particularly large linewidth and features a Fano profile, which is likely the consequence of the magnetic exchange coupled to the vibrations of the Co atoms. Interestingly, the A$_{1g}$ mode linewidth decreases below $T_c$, whereas the linewidth of the B$_{1g}$ mode increases. The DFT calculations generally agree with the measured phonon frequencies. However, fine frequency differences in the two phases cannot be correctly predicted since the DFT calculations do not account for the spin fluctuation effects.

\ack

We gratefully acknowledge discussions with R. Hackl.
This work was supported by the Serbian Ministry of Education, Science and Technological Development under Projects ON171032, III45018 and ON171017, by the European Commission under H2020 project VI-SEEM, Grant No. 675121, as well as by the DAAD through the bilateral Serbian-German project (PPP Serbien, grant-no. 56267076) "Interplay of Fe-vacancy ordering and spin fluctuations in iron-based high temperature superconductors." Work at Brookhaven is supported by the U.S. DOE under Contract No. DE-SC0012704 and in part by the Center for Emergent Superconductivity, an Energy Frontier Research Center funded by the U.S. DOE, Office for Basic Energy Science (C.P.). Numerical simulations were run on the PARADOX supercomputing facility at the Scientific Computing Laboratory of the Institute of Physics Belgrade. M.M.R also acknowledges the support by the Deutsche Forschungsgemeinschaft through Transregio TRR 80 and Research Unit FOR 1346.

\bibliographystyle{unsrt}

\end{document}